# Individual Privacy vs Population Privacy:
# Learning to Attack Anonymization

Graham Cormode

October 12, 2018

## Abstract

Over the last decade there have been great strides made in developing techniques to compute functions privately. In particular, Differential Privacy gives strong promises about conclusions that can be drawn about an individual. In contrast, various syntactic methods for providing privacy (criteria such as $k$-anonymity and $l$-diversity) have been criticized for still allowing private information of an individual to be inferred.

In this report, we consider the ability of an attacker to use data meeting privacy definitions to build an accurate classifier. We demonstrate that even under Differential Privacy, such classifiers can be used to accurately infer "private" attributes in realistic data. We compare this to similar approaches for inference-based attacks on other forms of anonymized data. We place these attacks on the same scale, and observe that the accuracy of inference of private attributes for Differentially Private data and $l$-diverse data can be quite similar.

## 1 Introduction

Anonymization and privacy has occupied the computer science research community for over a decade now (with efforts in statistics going back further). Initial efforts focused on trying to modify microdata, via reducing the precision of data (coarsening values, forming tuples into groups). These generally tried to achieve syntactic requirements: famously, $k$-anonymization [17], and subsequently an alphabet soup of further definitions ($l$-diversity [15], $t$-closeness [14], and so on). However, various attacks have been shown on these models which reduced confidence in their suitability for data release. In this case, a successful attack has been defined (de facto) as an analysis of the released data which allows an observer to make guesses as to a particular attribute of a tuple which are correct with probability larger than is intended by the anonymizer [19, 8, 13].

In parallel, a more principled approach to privacy has arisen, in the form of differential privacy [6]. In its simplest form, differential privacy releases statistics, by computing the exact value of the statistic, and then adding random noise. The random noise is chosen so that the influence of any individual on the statistic is masked by the noise. This leads to the observation that it is safe to provide one's personal data to a mechanism implementing differential privacy, since the conclusions that can be drawn are (broadly) the same, irrespective of whether or not one's data is collected. Consequently, differential privacy has started to be adopted as a 'gold standard' of privacy, and is becoming widely used.

In this paper, we study an attack on differential privacy, in the sense described above: it is possible to infer information about an individual with non-trivial accuracy. This does not violate any of the (proven) properties of differential privacy. It leads to the seeming contradiction that regardless of whether an individual contributes or withholds their data to a study (or reports false data), we can nevertheless use observable information about them to infer their private information. The attack is ultimately quite simple, and can be executed with minimal computational effort.

# 2 Differential Privacy

## 2.1 The Model

We adopt the canonical model of data from the initial work on anonymization: the input is a table of tuples $T$, and each tuple $t \in T$ corresponds to an individual. There is one distinguished attribute which is considered 'sensitive' (stereotypically, a disease suffered by the individual, or their salary band, say). The remaining attributes are considered partially identifying: demographic information about the individual, such as their age, gender, approximate location, ethnicity and so on. These are referred to in the literature as quasi-identifiers. Separately, each attribute value applies to a large number of individuals in the data, but taken collectively they are often sufficient to uniquely identify an individual [10].

For syntactic approaches to anonymization, the objective is to modify the input data by various operations so that it achieves a given property. The modifications may be to coarsen attributes (replace exact birth date by only year of birth), suppress some attributes entirely, or form tuples into groups and separate the multiset of sensitive values corresponding to each group. In most prior work, the effect of the data modification results in groups of tuples. The goal may be to ensure that each group has at least $k$ members ($k$ anonymity [17]); each group has at most a $1/l$ fraction of tuples with the same sensitive attribute value (simplified $l$-diversity [15]); and so on.

The differential privacy approach is different, in that it does not explicitly publish microdata. Instead, a query is posed on the data to the data owner. The data owner computes the exact answer of the query, then perturbs the answer by adding appropriate statistical noise. We focus on the core case when all queries are simply a collection of count queries: to count the number of individuals in the data who satisfy certain predicates. Further, we consider the case where there is only a single round of querying (asking for various counts): there is no adaptive querying in response to previous results. In the literature, these are known as histogram queries or contingency tables and the formula for amount of noise is straightforward [2]. Note that it is trivial to implement this in a publishing model, when the data owner chooses a collection of statistics to publish, and applies the differential privacy mechanism.

For a given query $q$, we must compute its *sensitivity* $s_q$, which is the maximum influence that any individual can have on the answer vector. The influence is measured under the $L_1$ norm. Given a parameter $\epsilon$, the data owner computes the vector of answers and adds iid noise to each entry. This noise may be drawn from a (continuous) Laplacian with parameter $s_q/\epsilon$, or from a (discrete) symmetric geometric distribution with parameter $\alpha = \exp(-\epsilon/s_q)$ [9]. This guarantees that, for any individual, the probability of any property holding on the output of the algorithm is within a multiplicative factor of at most $\exp(\epsilon)$ of the same property holding on the dataset with that individual's data omitted. Less formally, the intent of this definition is that even with substantial knowledge about other individuals in the data, it is still hard to deduce any property of a targeted individual.

Dalenius provided an oft-quoted definition of disclosure resulting from data release: "If the release of the statistic $S$ makes it possible to determine the (microdata) value more accurately than without access to $S$, a disclosure has taken place..." [5]. Following this definition, we define an *attack* on anonymized data as, given a target individual, to try to discover their sensitive value. More precisely, given access to the output of the anonymization process, and some knowledge about a targeted individual such as their easily observable characteristics (age, sex, ethnicity: quasi-identifiers), to infer their associated sensitive value.

This has been formalized in the literature as attribute (non-)privacy [12]: This definition applies to the earliest linking attack of Sweeney on de-identified data [18]; the homogeneity attack on $k$-anonymous data [15]; the minimality attack, which conditions on knowledge of the algorithm used to create the anonymized data [19]; and the deFinetti attack (discussed in greater detail in Section 3) [13]. All these attacks demonstrate disclosure, in the sense used by Dalenius. The success of the attack may be characterized by its accuracy: the fraction of input tuples for which the attack gives the correct value. Clearly, this probability will vary as a function of the data, and of parameters of the anonymization.

## 2.2 Accurate Differentially Private Classifiers

In line with previous attacks on privacy [13, 3], we describe a method to build an accurate classifier which, given the quasi-identifier of an individual, predicts their sensitive attribute. For simplicity, we initially assume that all attributes in the data are categoric and of relatively low cardinality, and later discuss this assumption.

Many classifiers have been proposed in the machine learning literature, but the simplest is the Naive Bayes classifier (NB). We will build an NB classifier which aims to predict the SA value given the evidence of QI values of an individual. On an unanonymized data table $T$, the classifier is built by computing the conditional distributions of each attribute given the target value $s$, i.e. $\Pr[t_i|s]$ for each of the $m$ QI values, $t_i$. The prediction for the tuple $t$ of QI values is given by[1]

$$\hat{s}(t) = \arg \max_{s \in SA} \Pr[s] \prod_{i=1}^{m} \Pr[t_i|s]$$

Observe that the parameters of this classifier are easy to learn: given a tuple $r$, $r_i$ is its $i$th component, and $r_s$ is its sensitive value. Then we use

$$\Pr[t_i|s] = \frac{\Pr[t_i \cap s]}{\Pr[s]} \approx \frac{|\{r \in T : r_i = t_i \cap r_s = s\}|}{|\{r \in T : r_s = s\}|}$$

That is, to build the classifier, we need the counts of values present in $T^i$, the $i$th column of table $T$:

$$\forall s \in SA, 1 \leq i \leq m, v \in T^i : |\{r \in T : r_i = v \cap r_s = s\}| \quad (1)$$

$$\text{and} \quad \forall s \in SA : |\{r \in T : r_s = s\}|$$

from which we can derive the required conditional probabilities.[2]

Observe that the sensitivity of this query is actually quite low: although there are a moderate number of parameters, the influence that any tuple can have on the answer vector is $m + 1$: each individual is counted in at most $m$ joint distributions, plus the marginal distribution of SA values. This is exactly the approach proposed in work on releasing histograms and contingency tables [2]. Thus the attack proceeds by requesting all counts listed in (1). Some small corrections are required: first, the noise may cause some counts to become negative. These can simply be adjusted up to the smallest feasible value, i.e. 0. Secondly, due to the noise, we do not have that the sum of the counts for attribute $i$ with SA value $s$ is certain to equal the marginal count of $s$. We remedy this by simply defining

$$\Pr[t_i|s] := \frac{1 + \max(0, |\{r \in T : r_i = t_i \cap r_s = s\}|)}{\sum_{t \in T_i} 1 + \max(0, |\{r \in T : r_i = t \cap r_s = s\}|)}.$$

Here, the addition of 1 is the standard Laplacian correction. Further, we can use

$$\Pr[s] = \frac{\sum_{i=1}^{m} \sum_{t \in T_i} 1 + \max(0, |\{r \in T : r_i = t \cap r_s = s\}|)}{\sum_{s' \in SA} \sum_{i=1}^{m} \sum_{t \in T_i} 1 + \max(0, |\{r \in T : r_i = t \cap r_s = s'\}|)}$$

and therefore reduce the sensitivity of the query from $m + 1$ to $m$.

This "naive attack" is therefore trying to emulate the corresponding classifier built with exact counts. So we do not expect the attack to predict values better than this noiseless classifier. Rather, the question is to what extent does the noise introduced by differential privacy degrade the accuracy of the classifier. Historically, Naive Bayes has been shown to get tolerable accuracy (when compared to other classification methods) with a moderate number of attributes, in the range $m = 3$–$10$, say. The sensitivity, also $m$, is therefore quite low: for settings of $\epsilon$ in the range 1.0 to 0.1, the absolute value of the noise introduced to counts is in the single to double digit range. While this is enough to mask the contribution of any individual, the crucial issue is that for a large enough dataset ($|T|$), the effect on the derived conditional probabilities is still relatively small, and thus the classifier learns approximately the same correlations. Therefore, even while a single individual is dominated by the noise, the noise is in turn dominated by the signal emerging from the whole population.

---

[1]This can be normalized so that assignment over different $s$ values is a probability distribution.

[2]More correctly, we can draw the parameters for the classifier from a Dirichlet distribution given the observed counts.

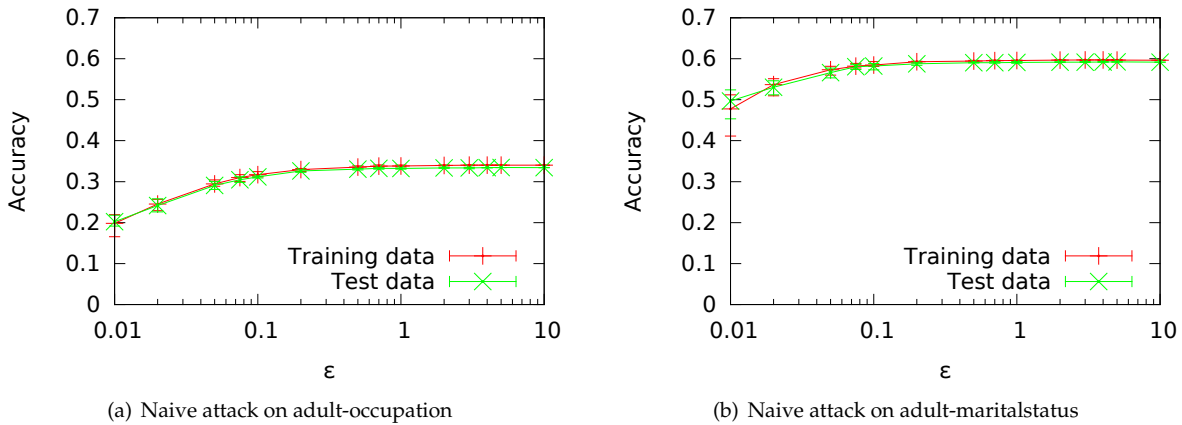

Figure 1: Naive attack on adult data set

## 2.3 Experimental Study

We implemented a proof-of-concept version of the "naive attack" in Python. We pick two data sets containing demographic data: the 'Adult' and 'Internet' data sets from the UCI Machine Learning repository [1]. The adult data set contains multiple attributes from a census-like survey of adults, while Internet contains details of a survey on Internet usage in 1997. From the adult data set, we selected the attributes workclass, level of education completed, occupation, sex, hours worked per week, and the binary attribute on whether income exceeds $50K. We bucketized hours worked per week into [0–25], [26–40], [41–60] and 60+. After removing tuples with missing sensitive values, there are a total of 30718 tuples in the adult data set. For the Internet data set, we selected the attributes age (treated as categoric but not bucketized), country, education level, gender, occupation, marital status and household income. Income is provided bucketed into ranges $10-20K, $20-30K, $30-40K, $40-50K, $50-75K, $75-100K, over $100K, below $10K, and 'not given'. There are a total of 10108 tuples in the Internet data set. In neither case did we remove tuples with missing QI values, since the classifier is relatively robust to their presence.

Our classifier was implemented to generate conditional probabilities given counts with geometric noise added as a function of the parameter $\epsilon$. We evaluated the accuracy of the classifier by then providing the quasi-identifiers of each input tuple in turn, and counting the fraction for which the correct sensitive attribute was guessed (based on taking the value deemed most likely by the classifier).

Figure 1 shows the accuracy of the naive attack (fraction of correct predictions) as $\epsilon$ is varied. We plot the accuracy when the sensitive attribute is 'occupation' (Figure 1(a)) and when it is 'marital status' (the same attributes are targeted in prior work on attacks [3, 13]). Note that there are fourteen categories of occupation in the data, and the trivial method which simply predicts the most frequent occupation achieves 13% accuracy. There are seven categories of marital status, and the most common occurs 45% of the time. The classifier is quite successful at learning the target attribute: the plot shows the minimum, maximum and mean accuracy across nine independent iterations (over different draws of the noise on the count queries). When $\epsilon$ is high, effectively no noise is being added, so the rightmost points are essentially the result of applying this classifier on the original data. As $\epsilon$ is decreased, the accuracy reduces gradually, despite more noise being added. Prior work on differential privacy has used values of $\epsilon$ in the range $\epsilon = 0.1\dots 1$. Even at $\epsilon = 0.01$, the attack achieves a non-trivial accuracy.

The plots in Figure 1(a) show the accuracy evaluated over two data sets. On the 'training data', we train the classifier on (noisy) counts, and then evaluate its accuracy on the same training data. On the 'test data', we evaluate the same classifier on a withheld test data set. This shows that the classifier is about as accurate on the withheld data as on the training data.

Figure 2 shows a similar experiment on the Internet usage data set, for the targets of household income and occupation (Figures 2(a) and 2(b) respectively). Income has nine distinct values, the most frequent of which occurs 18% of the time; Occupation has five distinct values, and the most frequent occurs 23% of the

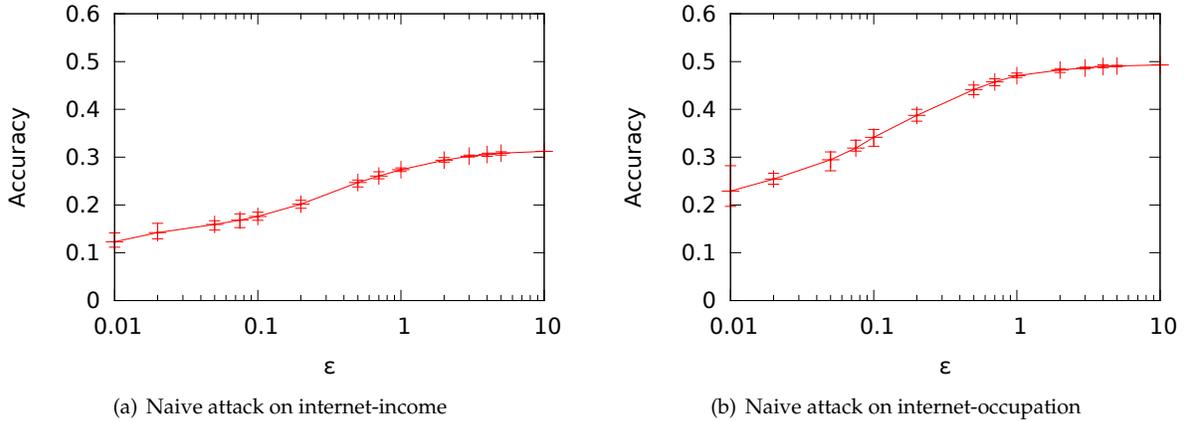

Figure 2: Naive attack on Internet data set

(a) Naive attack on internet-income  
(b) Naive attack on internet-occupation

time. There is a similar trend: the noiseless classifier is quite accurate, and increasing the amount of noise gradually reduces this accuracy.

Note that the classifier's output on each tuple can be interpreted as a probability distribution over sensitive values. Thus we can focus on those tuples for which the classifier places more of its weight on a particular outcome. Figure 3 shows the accuracy (summarizing nine separate, independent repetitions) when restricted to only those tuples for which the classifier's belief is greater than 0.8. They show that on these tuples, the classifier is indeed quite accurate. In the adult data set, there are a few hundred such tuples on average (around 1% of the data) when the target is occupation, and over 8000 (25%) for marital status. For occupation on the Internet data, there are around 1700 high confidence tuples (17%), and 300 (3%) for income. In other words, there is an appreciable minority of tuples on which the classifier is more confident, and justifiably so: reaching 85% accuracy on this subset in some cases.

Note that the classifier is quite lightweight to implement. The time to load the data, compute the required counts, add the noise, and apply the classifier to the collection of QI values was about 1 second on the Internet data (10K tuples) and 3 seconds on the adult dataset (32K tuples).

## 2.4 Discussion

The experiments show that this "attack" can be quite effective: we can learn supposedly private information of an individual with reasonable accuracy. In other words, even under differential privacy, disclosure can take place, in the sense used by Dalenius. Does this contradict the claims for differential privacy? No, in fact it is quite in line with the guarantees of differential privacy. These promise (informally) that what we can learn from the differentially private data is broadly the same, whether or not any individual contributes their true information, false information, or withholds their information entirely.

This is respected by the attack: rather than directly learning properties of an individual, it is learning properties of a population. The model we learn of the whole population is largely unaffected by any individual's data. However, the potential privacy issue is that this population model is quite accurate at predicting private information at the individual level, given an honest cooperative majority. This is seen most clearly in the experimental study: the classifier is about as accurate on the 'test set'. The test set contains individuals who never contributed any information about themselves to building the classifier, yet because they sufficiently resemble the training set, we can still predict their private information.

In fact, this consequence is in some ways anticipated by previous negative results on the possibility of privacy against sufficiently powerful attackers. Dwork [7] illustrates this with the example of an attacker who knows that "Terry Gross is two inches shorter than the average Lithuanian woman": the attacker can learn the average height of the Lithuanian population with high accuracy from a large survey of Lithuanians, and hence learn the individual height of Terry Gross. This example is perhaps easy to dismiss, due to the relative innocuousness of learning anyone's height, and the apparent absurdity of somehow knowing

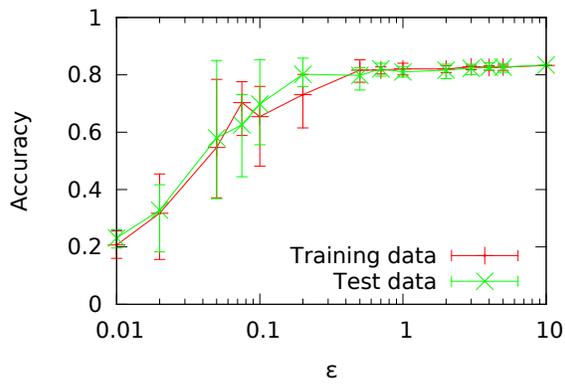
(a) Naive attack with high confidence on adult-occupation

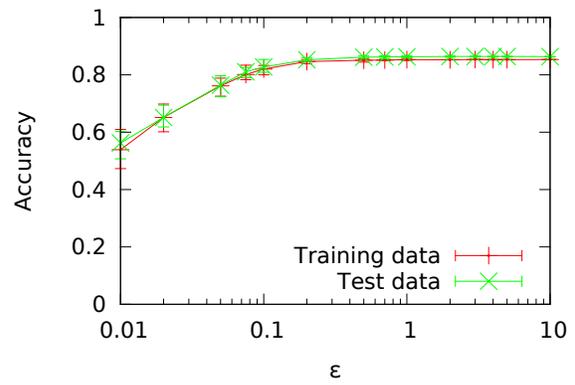
(b) Naive attack with high confidence on adult-maritalstatus

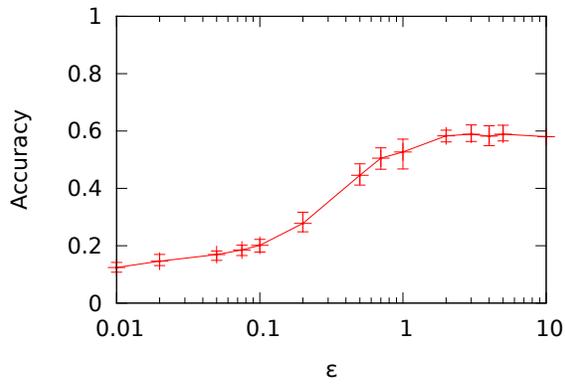
(c) Naive attack with high confidence on internet-income

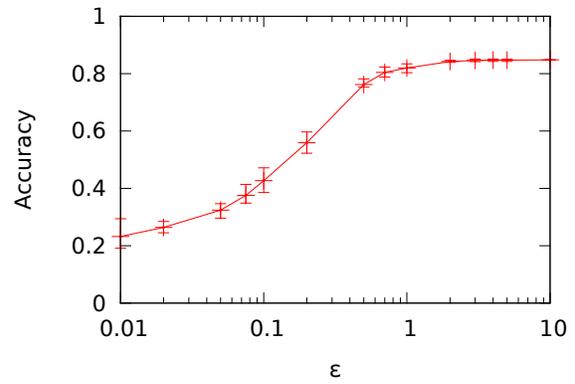
(d) Naive attack with high confidence on internet-occupation

Figure 3: Naive attack restricted to high confidence predictions

the connection between an individual's height and that of an East European nation. Yet, in our setting, if we replace this knowledge by the assumption that "Terry Gross' (private) height is correlated with her (observable) age, sex, ethnicity etc., in much the same way as for the population in this survey", then the attack becomes much more plausible. The key differences are that the knowledge of an obscure fact is replaced with a reasonable assumption, and the correlation can be easily learned from data which meets the differential privacy requirement.

One may still question whether learning about an individual when they never reveal their private data to anyone is truly an attack in the spirit of the term. We argue that an attack is in the eye of the beholder: if an attacker now believes that they know an individual's private information with high confidence, then he too is indifferent as to whether the victim contributed to the study which enabled the inference. This is exactly captured by Dalenius' notion of disclosure: this inference would not have been possible without the release of information. Consider the case of a multi-year, many-participant, multi-million dollar genetic study: the attacker would not go to this effort to create the necessary classifier, but can build one from the released statistics for free. The core issue is that latent properties of a population, when learned, can compromise the privacy of an individual. This example has showed that this is not an academic concern, but can be instantiated easily and cheaply on typical data.

## 2.5 Variations

**Choice of Attributes.** In general, adding more attributes to a Naive Bayes classifier improves its accuracy. However, in our setting, each extra attribute increases the sensitivity of the query required, increasing the noise introduced. There is therefore a delicate tradeoff: the increase in classifier accuracy from the increased knowledge of correlations, against the loss in accuracy due to more noise. We leave more detailed discussion of this aspect for future study.

**Continuous attributes.** Naive Bayes is less effective when treating continuous attributes (such as height in millimetres) as categoric: there is too little evidence for any fixed value to accurately learn the underlying correlation. There are two natural fixes: the first is to model the distribution as, e.g., a Gaussian, and to learn the parameters of the distribution (mean, variance) via appropriate count queries (which remain low sensitivity). The second is to 'bucketize' the attribute based on some natural bucketing scheme to make it categoric: group height into 10cm intervals, for example. This places together values which are semantically similar, so that appropriate correlations can still be determined. The same approach works when an attribute is categoric but has very high cardinality: domain-aware bucketing or coarsening can ensure that each count is sufficiently large to be unaffected by the noise. For example, location data at the street level might be coarsened to the town, state or country level, depending on the detail of the data. We assume that the adversary is able to request data according to the bucketing of their choice.

**Other Classifier Choices.** We focus on Naive Bayes due to its simplicity, the relatively small number of parameters and the minimal modeling skill needed in its use. Moreover, it is successful in this case because it has a low sensitivity. Naturally, one can also consider other classifiers. A first step is to consider higher-degree correlations: we can remove independence assumptions, e.g., by replacing separate factors of sex and age-range with the joint distribution of sex and age-range conditioned on the SA. This further reduces the sensitivity of the query (and hence the noise), but needs to be approached with care: just as in traditional Naive Bayes, partitioning up the space may lead to too few examples in each category. In our setting, this is further worsened since it increases the relative amount of noise added to each count.

More generally, other classifiers can be built. Indeed, there has been much prior work discussing the suitability of differential privacy for accurate learning [16, 11].[3] The chief concern is to consider the sensitivity of the classifier: extending to Bayesian networks does not substantially alter the sensitivity, which depends only on the number of nodes in the network. Various regression models also have low sensitivity. On the other hand, building a large number of random decision trees has high sensitivity, and requires the data owner to respond interactively to many queries as the classifier determines where to split each node.

---
[3]In particular, McSherry and Mironov point to the potential for accurate personalized predictions to compromise the privacy of an individual [16].

# 3 Comparison to the deFinetti Attack

The attack described is inspired by the deFinetti attack on syntactically anonymized data, introduced by Kifer [13]. The goal of the deFinetti attack is to build a classifier that, given the quasi-identifiers of a tuple in a group, is able to predict the corresponding sensitive attribute value. In other words, the approach and goal of the deFinetti attack is the same as the above naive attack. Moreover, the instantiation of the attack was also done based on Naive Bayes as the classifier, and the same dataset, meaning that we can compare the methods in more detail.

## 3.1 The deFinetti Attack

We now briefly summarize the attack. As in the previous study, we fix the method used to anonymize data as Anatomy, and the classifier as Naive Bayes, with the understanding that both of these can be replaced by other methods.

Anatomy is a grouping-based anonymization method [20]. Given a parameter $l$, it partitions the input data of size $n$ into groups of size $l$ (or $l+1$ when $l$ does not divide $n$), such that in each group there is at most one tuple for any given SA value. The published version of the data therefore includes the multiset of QIs in each group, and the multiset of SAs in each group, but withholds the exact mapping between them. The intent is therefore that a simplistic adversary should believe that any given QI in the group has a uniform chance of being associated with any particular SA in the same group, and thus bounds their belief in any association by at most $1/l$.

Kifer's observation is that there is sufficient correlation in the published group data that a more sophisticated adversary can learn a better set of beliefs (in the form of a classifier). The attack proceeds by guessing an initial random permutation for each group to map each QI value to one SA value. This induces a set of conditional distributions over each quasi-identifier value given a sensitive attribute value, which describes a Naive Bayes classifier. Indeed, this view is present in prior work: Brickell and Shmatikov describe essentially this first step in an earlier paper [3].

The deFinetti attack goes further by using the classifier (built from global information) to assess the relative likelihood of the current permutation in each group locally. One can then sample other possible permutations, and decide to adopt them in place of the current permutation: deterministically so if they are deemed more likely, and randomly so if they are less likely The likelihoods are determined by the current classifier. By iterating this process over many steps, the expectation is that the process converges to an accurate classifier.

The classifier can be used in conjunction with information from the anonymized data to make predictions. As originally described, given a QI value within a specific group, its predicted SA value can be that assigned to it based on the current permutation (or a history of recent assignments). This is the version described by Kifer [13], which we refer to as the 'permutation' method. Alternately, we can apply the classifier to the group, conditioned on the knowledge that the true SA is one of the $l$ possibilities. This we refer to as the 'group' method. Lastly, we can simply use the QI values to predict the SA value without explicit knowledge of the group structure, which we refer to as the 'open' method. Note that this last method should never be appreciably better than the NB classifier built from the original, noiseless data.

## 3.2 Experimental Study

The original study of the deFinetti attack presented experiments with relatively small values of $l$ (2,3,4) on the Adult data set. Starting from Kifer's original implementation, we extended the method to study its accuracy on larger group sizes, and under the additional accuracy measures (group and free, in addition to permutation). We describe our results for the same target, the 'occupation' attribute of the Adult data set. We perform a thousand iterations of the learning process: analyzing the accuracy of the classifier and the $L_1$ norm between subsequent joint distributions confirms that this is sufficiently many steps for the process to almost converge in most cases. We perform five repetitions of each experiments, and plot the max, min and average accuracy of each of the three methods to predict SA values (permutation, group and open).

Figure 4 shows the effects of the attack for different choices of the parameter $l$. Note that $l = 7$ is the upper limit, since there is one occupation with occurs just below 1/7 of the time. Figure 4(a) shows that

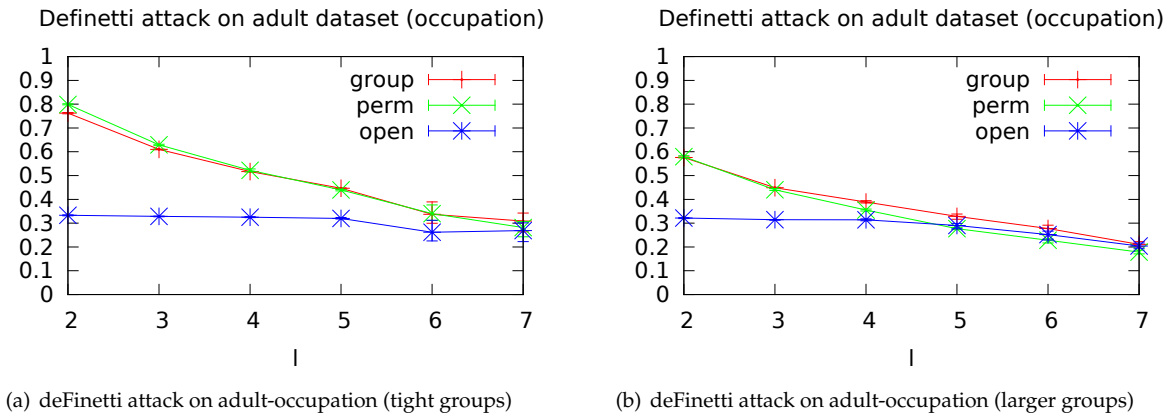

(a) deFinetti attack on adult-occupation (tight groups)  (b) deFinetti attack on adult-occupation (larger groups)

Figure 4: deFinetti attack on adult-occupation

our results agree with those reported in [13] for $l = 2, 3, 4$: in fact they show the attack to be more effective due to including tuples with missing QI values (giving more evidence), and choosing a slightly different set of QI attributes (giving a stronger classifier). The observation is that accuracy is clearly better than the trivial bound of $1/l$ when the adversary simply applies a uniform prior over all permutations from QI to SA values in each group, and does not attempt to learn a better classifier. Nevertheless, the trend is for the accuracy to decrease as $l$ increases.

The plot shows that the attack requires locating a target tuple within the published data in order to be truly effective. That is, without applying a constraint to the sensitive attribute, the 'open' approach, the accuracy is never better than the corresponding classifier trained on the true counts (see Figure 1(a), for a large $\epsilon$ value). Only when the attacker can guarantee finding the target tuple uniquely in the published data can he achieve the higher accuracy via the group or permutation method. For small groups, the permutation method (which conditions on other tuples in the same group) is slightly more accurate, but when the group grows large enough, the group method (which only predicts via the target tuple) becomes more accurate.

While the accuracy in all cases far exceeds $1/l$, at a group size of 7, the impact of the attack is less than that of the naive attack on differential privacy: the accuracy falls below 30%. This implies that the impact of the attack can be reduced by increasing group sizes. This is shown further in Figure 4(b). Here, we take the data output of Anatomy and merge together pairs of adjacent groups. The resulting output remains $l$-diverse, but groups now contain $2l$ tuples. Again, the accuracy exceeds $1/l$, but by a lower amount, and it falls below that of the noiseless classifier for $l = 5$ (group size 10). The permutation method is more clearly dominated by the group method, but both become close to the open method for larger groups. Increasing the size of the groups further decreases the ability of the classifier: when we merge together three groups (not shown), the permutation method degrades to about $1/l$ accuracy.

Similar results are seen on other data sets. Figure 5 shows the results on the Internet data set with 'income' as the SA value. Again, the open approach is never better than the noiseless classifier, and degrades as the group size increases. Being able to identify the target tuple within a group increases the accuracy above $1/l$, although this falls as the group size increases. Merging groups together further reduces the power of the attack: Figure 5(b) shows that the accuracy is the same as that of the noiseless classifier (see Figure 2(a)) when $l$ reaches 4. In larger groups (merging three groups into one), the accuracy is below $1/l$. As in the Differential Privacy case, we can focus in on those predictions in which the classifier places the highest confidence. We omit full details for brevity, but observe that there are similar trends: under the 'goup' definition, on small groups the attack has an appreciable number of individuals with high confidence predictions, which are indeed correct more often than for the rest of the population. As the group size is increased, the number of high confidence predictions dwindles, and the average accuract falls back to the global level.

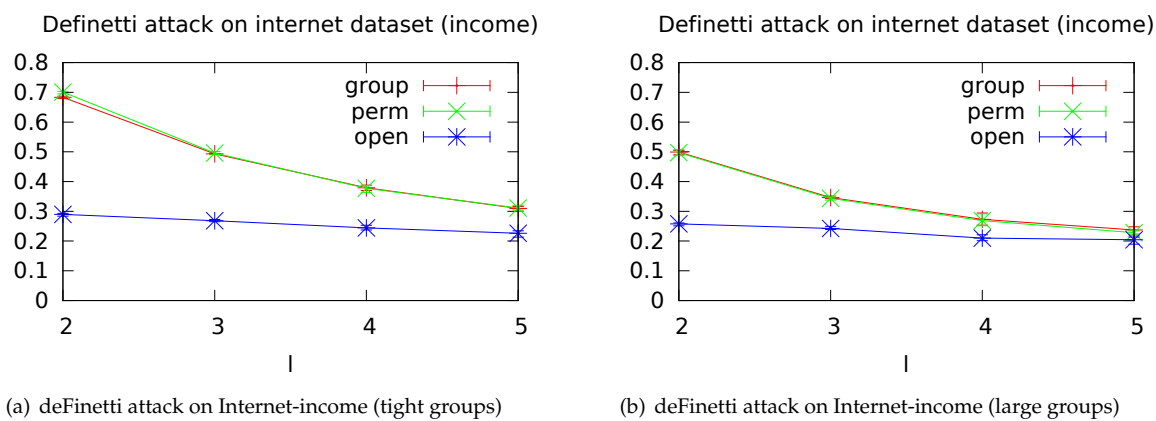

(a) deFinetti attack on Internet-income (tight groups)

(b) deFinetti attack on Internet-income (large groups)

Figure 5: deFinetti attack on Internet-income

## 3.3 deFinetti and Differential Privacy

Since its introduction, the deFinetti attack has been seen by some as the "last nail in the coffin" of syntactic privacy methods, and used to argue that only differentially private methods should be considered. Certainly, the attack can be potent when applied to data published with very small values the parameter $l$ and small groups. However, prior work on anonymization has typically used somewhat larger parameters: $l$ in the range 6-10 at least, and group sizes of 10-100 for higher security. In these situations, the attack loses its power: its accuracy reduces to $1/l$ (or worse), the same as simply guessing a value for each tuple uniformly from those in the group. Focusing only on privacy, it seems that the impact of the attack can be mitigated.

Moreover, the deFinetti attack is directly comparable to the above naive attack on differential privacy. Both model the data in the same way, target the same goal and attempt to build the same classifier to predict private information. Most of the effort in the deFinetti attack is expended in building the model of the population, which is trivial under differential privacy. Actually applying the classifier is then quite cheap in comparison.

The potential for the deFinetti attack to reveal more comes from additionally being able to condition on the limited set of possibilities within a group to eliminate or reduce the likelihood of possible values for a given tuple. That is, the 'open' method is (roughly) the accuracy from using properties of the population alone, and the additional power of group/perm comes from the additional information presented by the published data. This is impactful for small group sizes, but the advantage seems to be substantially diminished for larger groups sizes. For certain settings of parameters, differential privacy can be *more* susceptible to attack than the corresponding syntactic anonymization: the accuracy of the model of the population built in the latter case less than in the former case. Thus neither approach is immune to the possibility of attacking via building accurate classifiers.

Of course, syntactic anonymization is subject to other criticisms and attacks. But rejecting all such anonymizations because the deFinetti attack exists is erroneous: by the same logic, we should also abandon differential privacy as well. Rather, we need to consider more nuanced threat models: against what adversary do we require the anonymized data to withstand? This should differ depending on whether we are sharing anonymized data with a colleague in another department or releasing a data set publicly on the web. Depending on the perceived threats, and the consequences of a successful attack, it may be appropriate to use deidentification, syntactic privacy, differential privacy, or to withhold release entirely.

## 4 Discussion

The core issue highlighted here is that coarse properties of the population taken together quickly combine to build a model that can be applied to individuals with high accuracy. Existing anonymization methods tend to ignore this issue, and prior work has brushed this aside, assuming that such correlations can be ignored, or are known to all data users already. However, this is not sufficient: in reality, release of (anonymized) data may reveal hitherto unknown population parameters which compromise individual privacy. It seems unlikely that future efforts can satisfactorily resolve this issue, since in some settings, these populations statistics may represent exactly the desired *utility* of the data collection and publication. For example, the results of scientific studies are frequently parlayed to the public in the form of simple conditional probabilities ("drinking two glasses of wine each day reduces the chance of heart disease by 50%", say). Using such information for individuals to apply them to their own behavior, and modify it accordingly, is seen as a potential benefit of such studies. Meanwhile, the use of such information by a healthcare provider to prioritize patients for treatment may be seen as less desirable for the same individual.

To what extent should the attacks discussed in this paper chill our enthusiasm for differential privacy, or anonymization in general? In part, this depends on how severely we view the disclosure resulting from the attack. Here, the choice of terminology becomes an issue. The term "attack" has been inherited by the privacy world from the area of security. In security, an "attack" is typically understood to be a mechanism which compromises a supposedly secure system and permits access to the system or information within it. For example, a successful attack on an encrypted message typically reveals the cleartext content of the message to the attacker. Critically, our expectation is that a successful instance of an attack on security correctly indicates that it has indeed succeeded: the revealed message is determined to be the message that was sent.

For initial attacks on anonymization systems, it was clear that disclosure had occurred: Sweeney was sure that she had identified the health records belonging to the Governor of Massachusetts [18]. Subsequent attacks are more probabilistic in nature: the minimality attack (as introduced in [19] and analyzed in [4]) and the deFinetti attack [13] give an elevated belief in the association between a private value and an individual, but they do not provide certainty. Although we can empirically determine that these attacks are correct in associating an individual with their true value a large fraction of the time, they do not help to indicate with certainty for which individuals this is a true inference, and for which it is a false positive. The same is true of the naive attack we describe here: the classifier is often right, but we are never fully sure when.

For such probabilistic attacks, we must therefore decide what level of belief we can tolerate. This is naturally a function of the sensitivity of the information being inferred, and the way in which it is being used. For example, a blackmailer threatening to reveal a target's true sexuality to their family can perhaps tolerate making a few false accusations before finding a victim, while a law enforcement organization might require a much higher degree of suspicion before being granted a warrant to investigate further. Note however, as has been elaborated elsewhere, proceeding with an incorrect belief about an individual can be just as damaging to them as a correct belief (e.g. leading to denial of insurance coverage).

Clearly, there are many questions remaining over the use of data and its privacy-respecting dissemination. However, one message from this study is that it is not sufficient to naively apply differential privacy to data, and assume that this is sufficient to address all privacy concerns. Instead, much more careful deliberation about the consequences of the data release is needed.

## Acknowledgments

Thanks to Divesh Srivastava and Magda Procopiuc for detailed discussions, and to Dan Kifer for sharing his implementation of Anatomy and the deFinetti attack, and comments. Thanks also to Adam Smith, Entong Shen and Michael Hay for comments on earlier drafts.

## References

[1] A. Asuncion and D. Newman. UCI ML repository, 2007.

[2] B. Barak, K. Chaudhuri, C. Dwork, S. Kale, F. McSherry, and K. Talwar. Privacy, accuracy, and consistency too: a holistic solution to contingency table release. In *ACM Principles of Database Systems*, 2007.

[3] J. Brickell and V. Shmatikov. The cost of privacy: Destruction of data-mining utility in anonymized data publishing. In *ACM SIGKDD*, 2008.

[4] G. Cormode, N. Li, T. Li, and D. Srivastava. Minimizing minimality and maximizing utility: Analyzing method-based attacks on anonymized data. In *International Conference on Very Large Data Bases*, 2010.

[5] T. Dalenius. Towards a methodology for statistical disclosure control. *Statistik Tidskrift*, 15:429444, 1977.

[6] C. Dwork. Differential privacy. In *ICALP*, pages 1–12, 2006.

[7] C. Dwork. Differential privacy: A survey of results. In *Theory and Applications of Models of Computation*, 2008.

[8] S. R. Ganta, S. P. Kasiviswanathan, and A. Smith. Composition attacks and auxiliary information in data privacy. In *ACM SIGKDD*, 2008.

[9] A. Ghosh, T. Roughgarden, and M. Sundararajan. Universally utility-maximizing privacy mechanisms. In *ACM Symposium on Theory of Computing*, 2009.

[10] P. Golle. Revisiting the uniqueness of simple demographics in the us population. In *Workshop On Privacy In The Electronic Society*, 2006.

[11] S. P. Kasiviswanathan, H. K. Lee, K. Nissim, S. Raskhodnikova, and A. Smith. What can we learn privately? In *IEEE Conference on Foundations of Computer Science*, 2008.

[12] S. P. Kasiviswanathan, M. Rudelson, A. Smith, and J. Ullman. The price of privately releasing contingency tables and the spectra of random matrices with correlated rows. In *ACM Symposium on Theory of Computing*, 2010.

[13] D. Kifer. Attacks on privacy and deFinetti's theorem. In *ACM SIGMOD International Conference on Management of Data*, 2009.

[14] N. Li, T. Li, and S. Venkatasubramanian. $t$-closeness: Privacy beyond $k$-anonymity and $l$-diversity. In *IEEE International Conference on Data Engineering*, 2007.

[15] A. Machanavajjhala, J. Gehrke, D. Kifer, and M. Venkitasubramaniam. $\ell$-diversity: Privacy beyond $k$-anonymity. In *IEEE International Conference on Data Engineering*, 2006.

[16] F. McSherry and I. Mironov. Differentially private recommender systems: Building privacy into the netflix prize contenders. In *ACM SIGKDD*, 2009.

[17] P. Samarati and L. Sweeney. Protecting privacy when disclosing information: $k$-anonymity and its enforcement through generalization and suppression. Technical Report SRI-CSL-98-04, SRI, 1998.

[18] L. Sweeney. $k$-anonymity: a model for protecting privacy. *International Journal on Uncertainty, Fuzziness and Knowledge-based systems*, 10(5):557–570, 2002.

[19] R. C.-W. Wong, A. W.-C. Fu, K. Wang, and J. Pei. Minimality attack in privacy preserving data publishing. In *International Conference on Very Large Data Bases*, pages 543–554, 2007.

[20] X. Xiao and Y. Tao. Anatomy: Simple and effective privacy preservation. In *International Conference on Very Large Data Bases*, 2006.